\documentclass[apj,numberedappendix]{emulateapj}

\newcommand{\myemail}{kiyoto.yabe@nao.ac.jp}

\slugcomment{Accepted version...}

\shorttitle{Gas Inflow Rate and Outflow Rate}
\shortauthors{Yabe et al.}

\begin{document}


\title{The gas inflow and outflow rate in star-forming galaxies at $z\sim1.4$}


\author{Kiyoto Yabe\altaffilmark{1}, Kouji Ohta\altaffilmark{2}, Masayuki Akiyama\altaffilmark{3}, Fumihide Iwamuro\altaffilmark{2}, Naoyuki Tamura\altaffilmark{4}, Suraphong Yuma\altaffilmark{5}, Gavin Dalton\altaffilmark{6,7} \& Ian Lewis\altaffilmark{6}}

\altaffiltext{1}{Division of Optical and IR Astronomy, National Astronomical Observatory of Japan, 2-21-1, Osawa, Mitaka, 181-8588, Japan}
\altaffiltext{2}{Department of Astronomy, Kyoto University, Sakyo-ku, Kyoto, 606-8502, Japan}
\altaffiltext{3}{Astronomical Institute, Tohoku University, Aoba-ku, Sendai, 980-8578, Japan}
\altaffiltext{4}{Kavli Institute for the Physics and Mathematics of the Universe (WPI), The University of Tokyo, 5-1-5 Kashiwanoha, Kashiwa city, Chiba 277-8583, Japan}
\altaffiltext{5}{Institute for Cosmic Ray Research, The University of Tokyo, 5-1-5 Kashiwanoha, Kashiwa, 277-8582, Japan}
\altaffiltext{6}{Department of Astrophysics, University of Oxford, Keble Road, Oxford OX1 3RH, UK}
\altaffiltext{7}{STFC Rutherford Appleton Laboratory, Chilton, Didcot, Oxfordshire OX11 0QX, UK}

\email{\myemail}






\begin{abstract}
We try to constrain the gas inflow and outflow rate of star-forming galaxies at $z\sim1.4$ by employing a simple analytic model for the chemical evolution of galaxies. The sample is constructed based on a large near-infrared (NIR) spectroscopic sample observed with Subaru/FMOS. The gas-phase metallicity is measured from the [\ion{N}{2}]$\lambda$6584/H$\alpha$ emission line ratio and the gas mass is derived from the extinction corrected H$\alpha$ luminosity by assuming the Kennicutt-Schmidt law. We constrain the inflow and outflow rate from the least-$\chi^{2}$ fittings of the observed gas mass fraction, stellar mass, and metallicity with the analytic model. The joint $\chi^{2}$ fitting shows the best-fit inflow rate is $\sim1.8$ and the outflow rate  is $\sim0.6$ in unit of star-formation rate (SFR). By applying the same analysis to the previous studies at $z\sim0$ and $z\sim2.2$, it is shown that the both inflow rate and outflow rate decrease with decreasing redshift, which implies the higher activity of gas flow process at higher redshift. The decreasing trend of the inflow rate from $z\sim2.2$ to $z\sim0$ agrees with that  seen in the previous observational works with different methods, though the absolute value is generally larger than the previous works. The outflow rate and its evolution from $z\sim2.2$ to $z\sim0$ obtained in this work agree well with the independent estimations in the previous observational works.
\end{abstract}





\section{Introduction\label{Sec:Introduction}}
The gas flow into and out of galaxies have a significant impact on the formation and evolution of galaxies. The classical G-dwarf problem \citep[e.g.,][]{vandenBergh:1962p19415} can be naturally explained by the continuous inflow of primordial gas \citep[e.g.,][]{Larson:1972p28517}. More recently, it is found that the gas depletion timescale in local massive galaxies \citep[e.g.,][]{Wong:2002p28625} and high redshift galaxies \citep[e.g.,][]{Tacconi:2010p3235} is generally shorter than the build-up timescale of the stellar mass, which requires significant gas accretion \citep{Bouche:2010p12762}. The gas outflow from galaxies affects the interstellar medium (ISM) and the intergalactic medium (IGM). The outflow of enriched gas causes the decrease of the gas metallicity (hereafter metallicity) of galaxies and results in a presence of the stellar mass-metallicity relation of galaxies \citep[e.g.,][]{Tremonti:2004p4119,Peeples:2011p22365}. Feedback from outflows plays an important role for the star-formation activity and the structure formation of a galaxy \citep[e.g.,][]{Hopkins:2012p27908, Hopkins:2012p27897}. 

There is growing evidence to support the ubiquitous presence of the galactic scale outflow at high redshift. For instance, \citet{Weiner:2009p13953} found the significant outflow from DEEP2 star-forming galaxies at $z\sim1.4$, and also found that the outflow velocity and the absorption equivalent width increase with increasing stellar mass and star-formation rate (SFR). They also estimate the mass outflow rate from the inferred column density, characteristic size, and wind velocity and found that the outflow rate is comparable to the SFRs of the galaxies ranging from 10 to 100 M$_{\odot}$ yr$^{-1}$. Similar estimation of the outflow rate has been made at $z\sim0$ to $z\sim2$ \citep{Pettini:2000p13759, Steidel:2010p10312, Genzel:2011p27874, Martin:2012p27158,Bouche:2012p27878}.

On the other hand, the inflow of gas into galaxies is relatively difficult to observe, because weak metal absorption feature is expected in infalling gas with low density and low metallicity \citep[e.g.,][]{Fumagalli:2011p25055}, and the covering fraction may be very small \citep[less than 10\%, e.g.,][]{FaucherGiguere:2011p31200, Rubin:2012p27422}. The detectability might be very low unless a cold gas filament is exactly aligned with the line of sight \citep{Kimm:2011p31239,Goerdt:2012}. Various efforts have been made to detect the pristine gas accreting into galaxies by using absorption systems \citep[e.g.,][]{Fumagalli:2011p25058, Giavalisco:2011p28762, Stewart:2011a, Stewart:2011b, Bouche:2013p26633, Lehner:2013p26652, Fumagalli:2014}. In galaxies at $z\sim3$ observed with a NIR IFU spectrograph, \citet{Cresci:2010p17086} reported the positive metallicity gradient, which shows higher metallicity in the outer region rather than in the central region, suggesting that this is originated from very metal poor gas accreting into the center of the galaxies.

Recently, inflow and outflow rates of galaxies are constrained by cosmological simulations \citep[e.g.,][]{Keres:2005p26220} and analytic models based on the mass assembly history of dark matter halo \citep[][]{Bouche:2010p12762,Lilly:2013p26256}. \citet{Erb:2006p4143} show that a more simple chemical evolution model with moderately strong outflow well explains the distribution of gas mass fraction and metallicity of galaxies at $z\sim2$. With more detailed analysis of models employing both gas inflow and gas outflow, \citet{Erb:2008p4784} implies the presence of gas inflow at the rate twice larger than the SFR and gas outflow comparable to the SFR of the galaxy sample at $z\sim2$ by \citet{Erb:2006p4143}. Similar studies for galaxies at $z\sim3$ show that the closed box models with neither inflow nor outflow cannot explain the observations well \citep{Mannucci:2009p8028, Troncoso:2013p29493}.

The estimation of the gas inflow and outflow rate in this method, however, is still limited partly due to the large observational errors and the small sample size at high redshift. Recently, we construct a large NIR spectroscopic sample with significant H$\alpha$ detections from $\sim340$ star-forming galaxies at $z\sim1.4$ \citep{Yabe:2012, Yabe:2014} by using Fiber Multi Object Spectrograph (FMOS) \citep{Kimura:2010p11396} on the Subaru Telescope.  In this work, by using the simple analytical model of the chemical evolution, we try to constrain the gas inflow and outflow rate derived from the observed stellar mass, gas mass fraction, and the metallicity of this sample. Although the size of our sample is large, many [\ion{N}{2}]$\lambda6584$ emission lines, which are used for metallicity estimation, are weak, and we apply the stacking analysis. Hence, the number of the data points compared to the models is still limited. Under such circumstance, it would be more appropriate to use not the elaborate model with many free parameters but the simple analytic model.

Throughout this paper, we use the following cosmological parameters: $H_{0}=70$ km s$^{-1}$ Mpc$^{-1}$, $\Omega_{\Lambda}=0.7$, and $\Omega_{m}=0.3$. All magnitudes given in this paper are in the AB magnitude system.

\begin{figure}[!htb]
\includegraphics[angle=270, scale=0.50]{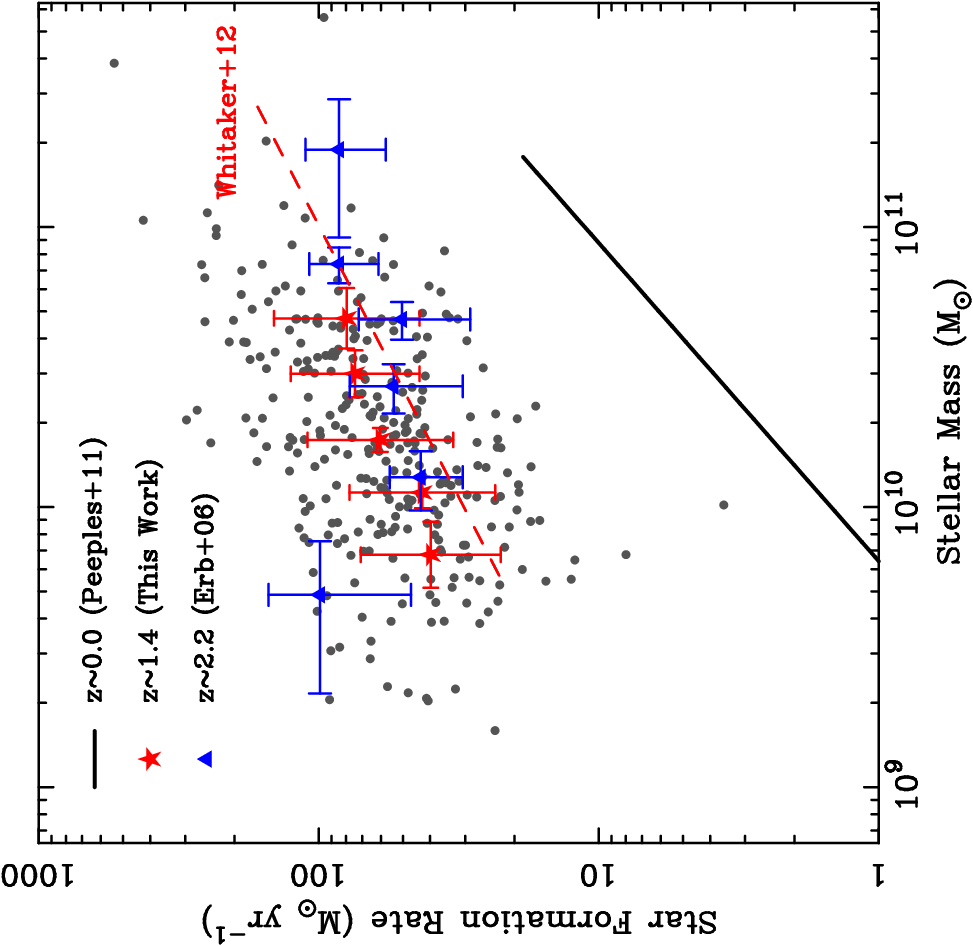}
\caption{The star formation rate (SFR) against the stellar mass of the sample. Objects of our sample at $z\sim1.4$ are shown by \textit{dots} and results from the stacking analysis are indicated by \textit{stars}. The sample at $z\sim0$ \citep{Peeples:2011p22365} and $z\sim2.2$ \citep{Erb:2006p4143} are indicated as a \textit{solid line} and \textit{triangles}, respectively. The result at $z\sim1.4$ in the previous study by \citet{Whitaker:2012p28816} is shown by a \textit{dashed line}. The stellar mass and SFR are scaled to the Salpeter IMF for consistency.\label{fig:MsSFR}}
\end{figure}

\section{Sample Selection}
\subsection{NIR Spectroscopic Sample at $z\sim1.4$}
The sample used in this work is originated from the $K$-band selected galaxy sample in the Subaru XMM-Newton Deep Survey and UKIDSS Ultra Deep Survey field (SXDS/UDS). The photometric redshifts (phot-$z$s) and the other parameters such as stellar mass are derived from the spectral energy distribution (SED) fitting thanks to the multi-wavelength photometric data covering from UV to mid-IR. The expected H$\alpha$ flux is also derived from the rest-frame UV luminosity density and the dust extinction calculated from the rest-frame UV color. From the sample, we selected galaxies with $K \le 23.9$ mag, phot-$z$ of $1.2 \le z_{ph} \le 1.6$, the stellar mass of $M_{*} \ge 10^{9.5}$ M$_{\odot}$, and the expected H$\alpha$ flux of $F(\textrm{H}\alpha) \ge 5\times10^{-17}$ erg s$^{-1}$ cm$^{-2}$. The details of the sample selection are described by \citet{Yabe:2012} and \citet{Yabe:2014}. The distribution of our sample on the diagram of stellar mass vs. SFR is presented in Fig. \ref{fig:MsSFR}. The target sample is consistent with the distribution of galaxies on the star-formation main sequence \citep[e.g.,][]{Daddi:2007p1460, Wuyts:2011p30356, Whitaker:2012p28816} within error bars.

We observed $\sim1200$ objects by using FMOS with low resolution mode covering wavelength range of $\sim0.9$ to $\sim1.8$ $\mu$m during the FMOS/GTOs, engineering runs, and open-use observations. The typical on-source exposure time is $3-4$ hours per one object. The basic data reductions were carried out by using the standard FMOS pipeline FIBRE-pac \citep{Iwamuro:2011p18754}. Multiple emission lines such as [\ion{N}{2}]$\lambda$$\lambda$6548, 6584, [\ion{O}{3}]$\lambda$$\lambda$4959, 5007, and H$\beta$ were detected in $\sim$35 \% of the resulting spectra. For objects with a single emission line, we take it as H$\alpha$ based on the phot-$z$ information, which is accurate enough with the uncertainty of $\sigma_{z}\sim0.05$ in this redshift range thanks to the multi-wavelength data in the SXDS/UDS. In total, significant H$\alpha$ emission lines with signal-to noise ratio (S/N) $\ge$ 3 are detected from 343 objects. The detected H$\alpha$ and [\ion{N}{2}]$\lambda$$\lambda$6548, 6584 emission lines are fitted with multiple Gaussian with free parameters of the redshift, line width, and the normalizations. For the detected [\ion{O}{3}]$\lambda$$\lambda$4959, 5007, and H$\beta$ lines, we also apply the same multiple Gaussian fitting. The effect of the mask in the FMOS OH-suppression system with low spectral resolution mode is taken into consideration during the spectral fitting process. The flux loss due to the fiber aperture effect is statistically corrected by using the size of the target; the typical correction factor is a factor of $\sim2$. The detailed fitting method and corrections are described by \citet{Yabe:2012}. The spectroscopic redshifts range from $z\sim1.2$ to $z\sim1.6$ with a median value of $z=1.410$, which agree with the phot-$z$s within $\sigma_{z}\sim0.05$.

The bright X-ray sources with $L_{X}\gtrsim10^{43}$ erg s$^{-1}$ are excluded from our original sample based on the catalogue by \citet{Ueda:2008p13177}. We also examine active galactic nucleus (AGN) contamination by using the [\ion{N}{2}]/H$\alpha$ and [\ion{O}{3}]/H$\beta$ line ratio diagnostics \citep[BPT;][]{Baldwin:1981p15095}. As possible AGN candidates, $\sim21$ objects are excluded from the H$\alpha$ detected sample. We also exclude objects with the calculated flux loss rate by the OH-mask is larger than 67 \% from the sample. By this threshold, the uncertainty of the reproduced flux is less than $\sim$10 \%. The details are described by \citet{Yabe:2012} and \citet{Yabe:2014}. In total, the number of the galaxy sample we use in this work is 271.

\def\arraystretch{1.5}
\begin{table*}
 \begin{center}
 \caption{Physical properties in five stellar mass bins}\label{tab:properties}
 \begin{tabular}{ccccc}
 \hline \hline
log($M_{*}$ [M$_{\odot}$]) & \# of galaxies & $\mu$$^{a}$ & $Z$ [Z$_{\odot}$]$^{b}$ & log(SFR [M$_{\odot}$ yr$^{-1}$]) \\
\hline
  9.83$\pm$0.12 & 55 & 0.66$\pm$0.05 & 0.61$\pm$0.05 & 1.60$\pm$0.25 \\
10.05$\pm$0.06 & 54 & 0.51$\pm$0.04 & 0.68$\pm$0.04 & 1.63$\pm$0.26 \\
10.24$\pm$0.04 & 54 & 0.48$\pm$0.05 & 0.66$\pm$0.04 & 1.78$\pm$0.26 \\
10.48$\pm$0.08 & 54 & 0.42$\pm$0.04 & 0.74$\pm$0.04 & 1.87$\pm$0.23 \\
10.67$\pm$0.11 & 54 & 0.31$\pm$0.03 & 0.86$\pm$0.03 & 1.90$\pm$0.26 \\
\hline
\end{tabular}
\end{center}
\begin{flushleft}
Notes. --- $^{a}$ Average gas mass fractions and standard errors are presented.
$^{b}$ The average metallicities are based on the stacking analysis, and the errors are estimated with the bootstrap resampling method.
\end{flushleft}
\end{table*}

\begin{figure*}[!htb]
\includegraphics[angle=270, scale=0.37]{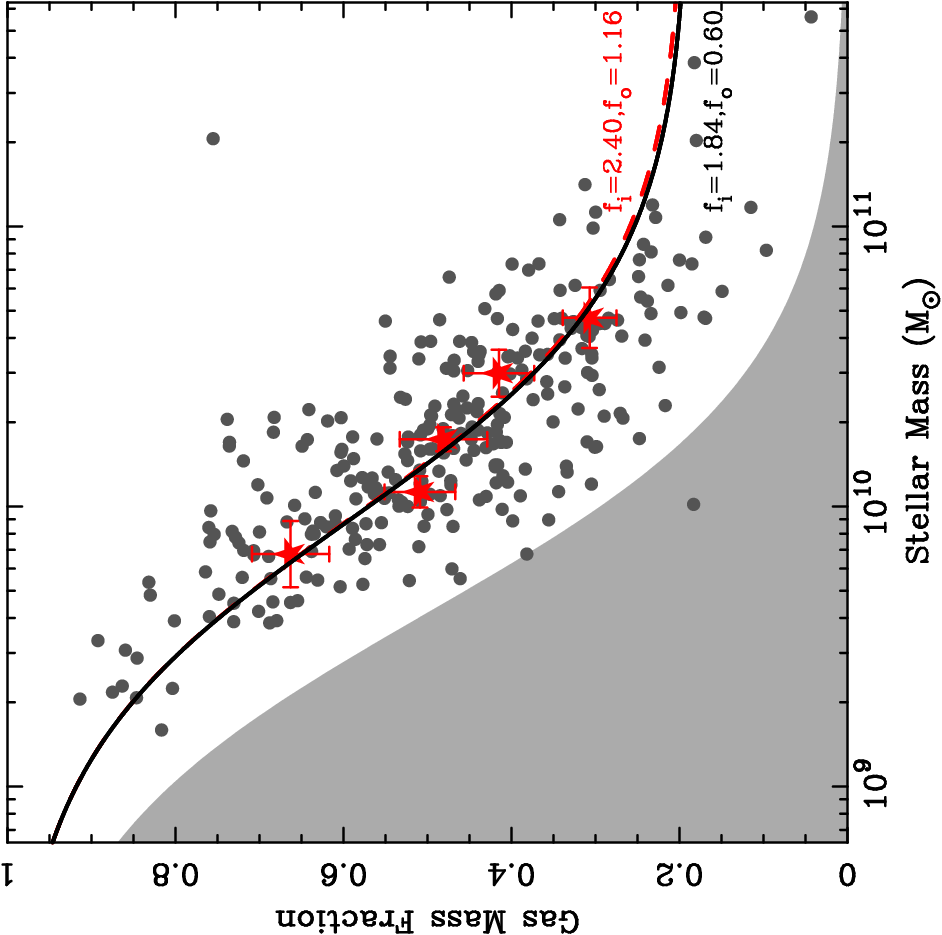}
\includegraphics[angle=270, scale=0.37]{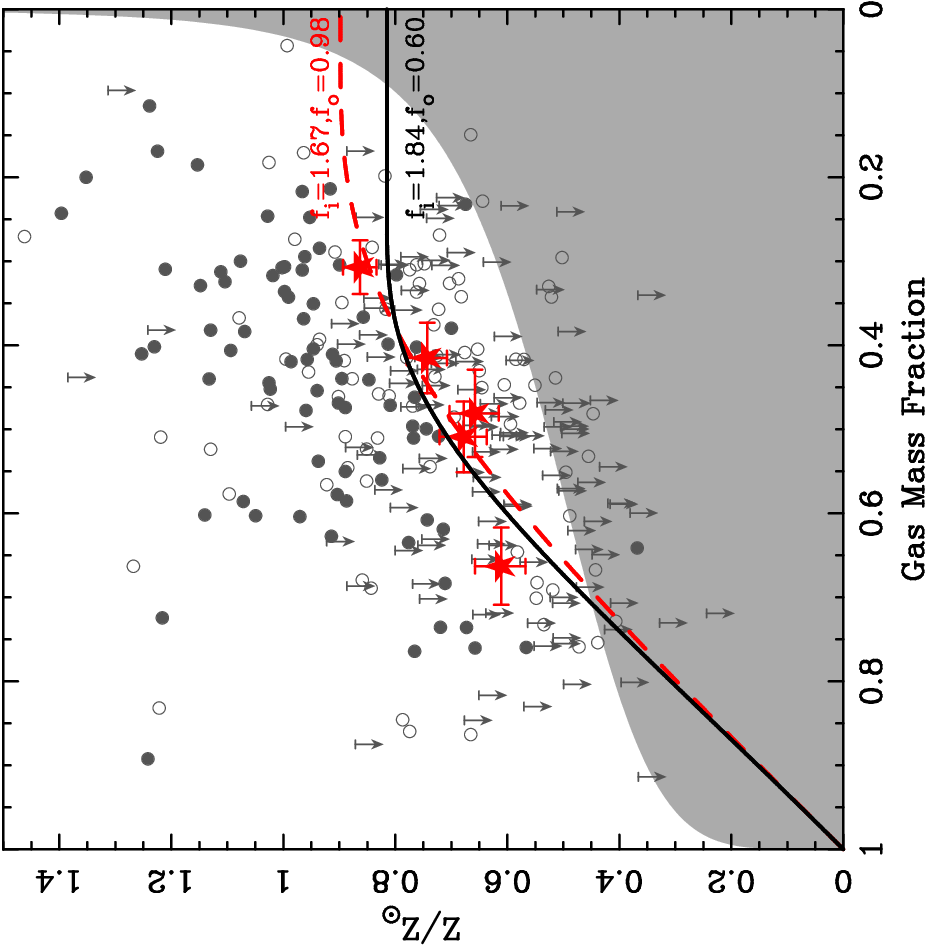}
\includegraphics[angle=270, scale=0.37]{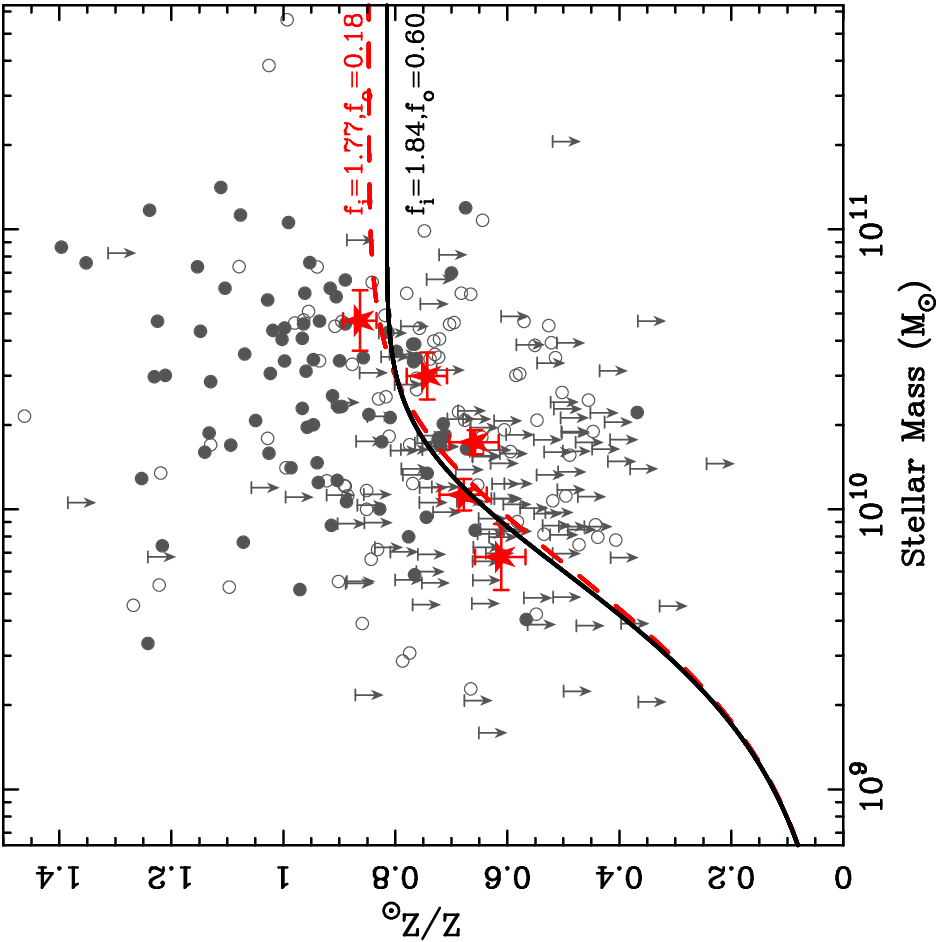}
\caption{The stellar mass vs. gas mass fraction ($M_{*}$ vs. $\mu$) diagram (\textit{left}), the gas mass fraction vs. metallicity ($\mu$ vs. $Z$) diagram (\textit{middle}), and stellar mass vs. metallicity ($M_{*}$ vs. $Z$) diagram (\textit{right}). Objects with [\ion{N}{2}]$\lambda$6584 lines with S/N $>$ 3.0 and 1.5 $<$ S/N $\le$ 3.0 are indicated by \textit{filled} and \textit{open} circles, respectively. Those with [\ion{N}{2}]$\lambda$6584 lines with S/N $<$ 1.5 are plotted as upper limits with values corresponding to 1.5$\sigma$. The average value in each stellar mass bin is presented by \textit{filled stars}. \textit{Red dashed lines} are the best-fit models by using the average data points in each panel, while \textit{black solid lines} are the best-fit models by the joint fitting. Observational limits for our sample are indicated by \textit{gray shaded regions} (see text for details). \label{fig:ObservedDataPoints}}
\end{figure*}

\subsection{Stellar Mass and Metallicity Measurement \label{sec:MassMetallicity}}
The stellar mass is calculated from the SED fitting by using the SEDfit code \citep{Sawicki:2012}. The detailed methods and assumptions are presented by \citet{Yabe:2012} and \citet{Yabe:2014}. Here, we fixed the redshifts of the objects to their spectroscopic redshifts. We use the Salpeter initial mass function (IMF), where the stellar mass is generally larger than that derived with the Chabrier IMF by $\sim1.8$ \citep[e.g.,][]{Erb:2006p4143}. In Section \ref{sec:CosmicEvolution}, we use this conversion factor for the fair comparison of the results to previous studies at other redshifts.

The gas-phase metallicity is derived from the [\ion{N}{2}]$\lambda6584$/H$\alpha$ emission line ratio. We use the calibration of N2 method by \citet{Pettini:2004p7356}. There exists large uncertainty in the calibration of the metallicity; the uncertainty of the N2 calibration itself is $\sim0.1$ dex \citep{Pettini:2004p7356} and that systematic uncertainty between various metallicity calibrations is $\sim1$ dex at most \citep{Kewley:2008p5141}. In Section \ref{sec:CosmicEvolution}, we use the conversion between the metallicity derived from the N2 calibration and those with other calibration by \citep{Kewley:2008p5141} in the comparison to the previous results. The resulting metallicity ranges widely from $\sim0.3$ to $\sim1.4$ Z$_{\odot}$. The distribution of the metallicity against the stellar mass is presented in the right panel of Fig. \ref{fig:ObservedDataPoints}.

As is already mentioned by \citet{Yabe:2012} and \citet{Yabe:2014}, a large part of our H$\alpha$ detected sample shows no significant [\ion{N}{2}] emission lines. We also apply the stacking analysis by dividing our sample into 5 stellar mass bins, in each of which the number of galaxies is $>50$. Here we use the stacking method by the obtained best-fit model spectra as a fiducial stacking method \citep[for details, see Section 3.2 of][]{Yabe:2014}. The obtained metallicity from the stacking analysis in each stellar mass bin is presented in the right panel of Fig. \ref{fig:ObservedDataPoints} as \textit{filled stars}.

\subsection{Gas Mass Fraction\label{sec:GasMassFraction}}
The gas mass is estimated from the observed H$\alpha$ luminosity by assuming the Kennicutt-Schmidt (K-S) law \citep{Kennicutt:1998p7470, Schmidt:1959p28292}. The derivation of gas mass by assuming the K-S law was also applied at high-redshift by \citet{Erb:2006p4143} at $z\sim2.2$. Although this relation is calibrated with local galaxies, various studies at high redshift show no strong redshift evolution up to $z\sim2$ \citep[e.g.,][]{Genzel:2010p29007}. The SFR surface density ($\Sigma_{SFR}$) is calculated from the intrinsic H$\alpha$ luminosity and the galaxy size. For the H$\alpha$ luminosity, we apply the correction of the dust extinction for nebular emission and the aperture effect of the FMOS fibers \citep[for details, see][]{Yabe:2012, Yabe:2014}. We use the conversion by \citet{Kennicutt:1998p7465} from the H$\alpha$ luminosity to the SFR. As the size of the galaxy, we use the $r_{50}$ derived from the image in $B$-band, which corresponds to the rest-frame wavelength of $\sim2000$ \AA\ and is expected to trace the global star-formation activity, after deconvolving by the typical seeing size. Although the regions traced by the rest-frame UV continuum may be different from those traced by H$\alpha$ emission, here we assume that both have the same size in this work. The obtained $\Sigma_{SFR}$ is converted to the gas (H$_{2}$ and \ion{H}{1}) surface density ($\Sigma_{gas}$) by using the K-S law with an index of $n=1.4$. Then, the gas mass is derived from $\Sigma_{gas}$ and the $r_{50}$. The obtained gas mass ranges from $\sim2\times10^{9}$ to $\sim1\times10^{11}$ M$_{\odot}$ with the average of $1.7\times 10^{10}$ M$_{\odot}$.

The obtained gas mass is compared to that derived from the dust mass assuming the gas-to-dust ratio. We estimate the average dust mass of our sample from the stacked far infrared (FIR) SED by using \textit{Herschel} Photodetector Array Camera \& Spectrometer (PACS) and Spectral and Photometric Imaging Receiver (SPIRE) data (Yabe et al. 2014, in prep.). The dust mass derived from the gray body fitting to the stacked SED is $1.2\times 10^{8}$ M$_{\odot}$. In the local Universe, it is known that the gas-to-dust ratio correlates with the metallicity well, i.e., galaxies with higher metallicity tend to show lower gas-to-dust ratio \citep[e.g.,][]{Leroy:2011p22814}. It is also reported that the dependence of the gas-to-dust ratio on the metallicity at $z=1-2$ is consistent with the local relation \citep[e.g.,][]{Magdis:2012p27223, Seko:2014}. From the metallicity dependent gas-to-dust ratio by \citet{Leroy:2011p22814}, we use the value of 150 in the metallicity range of our sample. Here, the gas mass in the calibration by \citet{Leroy:2011p22814} includes both H$_{2}$ and \ion{H}{1}. The gas mass derived from the dust mass and the gas-to-dust ratio is $1.9\times10^{10}$ M$_{\odot}$, which is well in agreement with the average value of the gas mass of $1.7\times 10^{10}$ M$_{\odot}$ derived by assuming the K-S law.

The gas mass fraction ($\mu$) is defined as $\mu=M_{gas}/(M_{*}+M_{gas})$, where $M_{gas}$ and $M_{*}$ are gas mass and stellar mass, respectively. The resulting gas mass fraction is widely distributed from $\sim0.2$ to $\sim0.8$ with the median (average) value of 0.47 (0.48). The gas mass fraction of our sample is compared with that obtained by using more direct CO measurements at similar redshift \citep{Tacconi:2010p3235,Daddi:2010p3611,Tacconi:2013p25494}. The mean gas mass fraction of our sample at stellar mass of $\sim10^{10.5}$ M$_{\odot}$ is $0.33$, which is in well agreement with that from the CO observations of $0.34$ in the same stellar mass range after the IMF conversion to Salpeter. We calculate the mean gas mass fraction of our sample in the 5 stellar mass bins, which are summarized in Tab. \ref{tab:properties}.

\begin{figure*}[!htb]
\includegraphics[angle=270, scale=1.00]{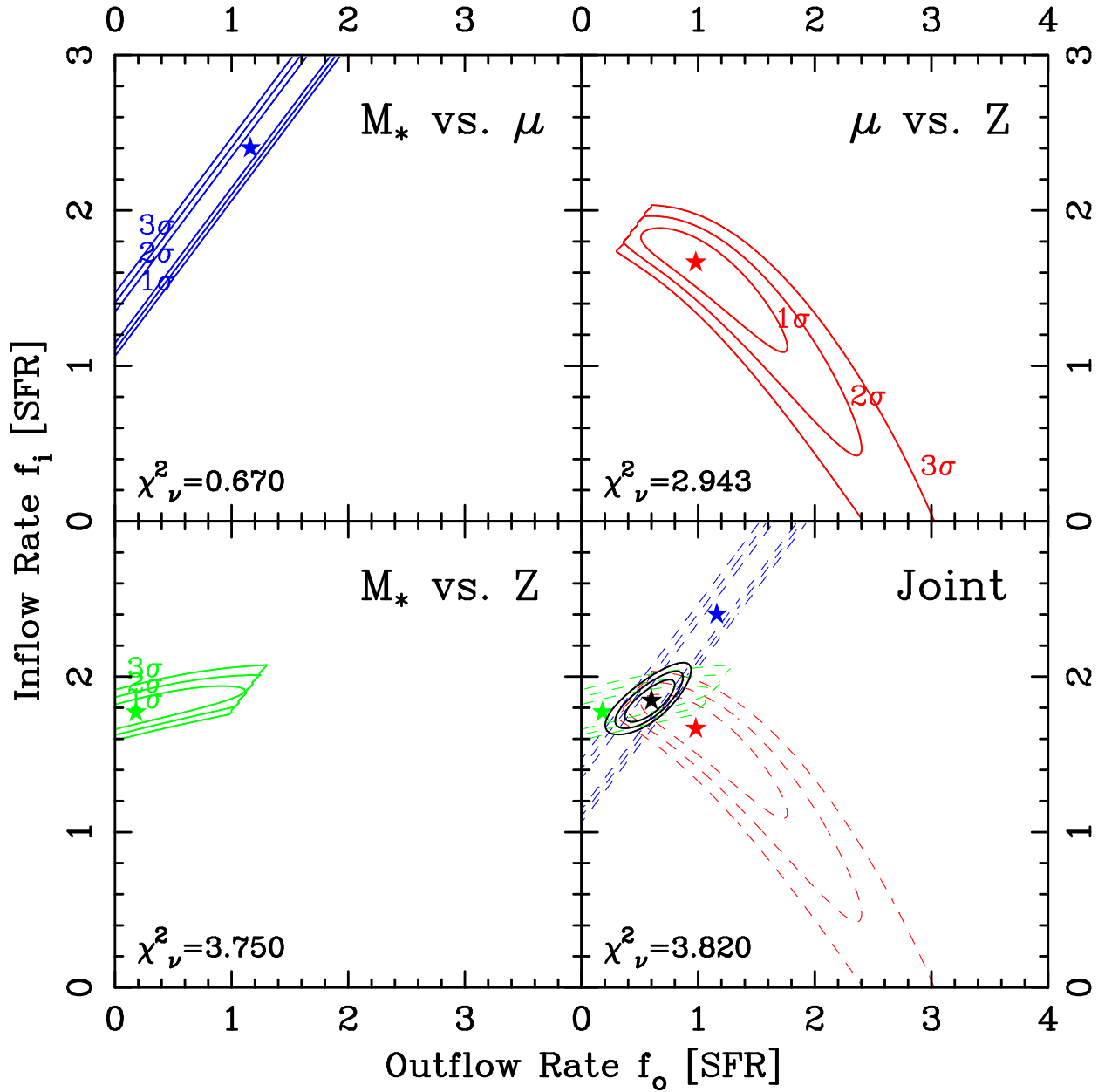}
\caption{$\chi^{2}$ contour maps for the sample at $z\sim1.4$ in the fitting of the stellar mass vs. gas mass fraction ($M_{*}$ vs. $\mu$) diagram (\textit{top left}), gas mass fraction vs. metallicity ($\mu$ vs. $Z$) diagram (\textit{top right}), stellar mass vs. metallicity ($M_{*}$ vs. $Z$) diagram (\textit{bottom left}), and the joint values (\textit{bottom right}). The best-fit values at the least $\chi^{2}$ are indicated by \textit{filled stars}. Contours are shown for 1$\sigma$, 2$\sigma$, and 3$\sigma$. \label{fig:Chisq:z14}}
\end{figure*}

\section{Results and Discussions}
\subsection{Relations Between Gas Mass Fraction, Metallicity, and Stellar Mass\label{sec:MsMu}}
In Fig. \ref{fig:ObservedDataPoints}, distributions of stellar mass, gas mass fraction, and metallicity are presented. Individual data points are shown by gray symbols including arrows as upper limits for metallicity and the average data points in five stellar mass bins are shown by filled stars. We use the metallicity derived from the stacking analysis as the average data points in the fitting described below. In the left panel, a tight anti-correlation between stellar mass and gas mass fraction can be seen; the gas mass fraction decreases with increasing stellar mass. In the middle panel, the metallicity increases with decreasing gas mass fraction. In the right panel, the mass-metallicity relation is shown, i.e., metallicity increases with increasing stellar mass.

Since the gas mass is derived from the intrinsic H$\alpha$ luminosity, which also correlates with the intrinsic SFR, the observable gas mass fraction is limited by the sample selection. Our sample at $z\sim1.4$ is selected with the expected H$\alpha$ flux limit of $5\times10^{-17}$ erg s$^{-1}$ cm$^{-2}$, corresponding to the SFR of $\sim10$ M$_{\odot}$ yr$^{-1}$. The gas mass is hence limited to $\sim5\times10^{9}$ M$_{\odot}$. In the the left panel of Fig. \ref{fig:ObservedDataPoints}, the corresponding limit on the stellar mass vs. gas mass fraction are presented as shaded region, which means that the shaded area cannot be observed due to our selection effect. In the middle panel, the limit of the gas mass fraction as a function of stellar mass is converted to a function of metallicity assuming the mass-metallicity relation at $z\sim1.4$ by \citet{Yabe:2014}. The existence of the selection effect has been pointed out in the sample of star-forming galaxies at $z\sim3$ by \citet{Mannucci:2009p8028}. They mentioned, however, that the deficit of galaxies in this region could be partly real by considering the strong correlation between stellar mass and gas mass fraction expected from theoretical models \citep[e.g.,][]{Dave:2011b}. 

In order to evaluate the selection effect on the gas mass fraction as a function of the stellar mass, we estimate the expected gas mass fraction of typical main-sequence galaxies at this redshift range. From the relation between stellar mass and SFR at $z\sim1.5$ by \citet{Wuyts:2011p30356}, we calculate the SFR at a given stellar mass, and then, the SFR is converted to the gas mass by assuming the K-S law. The estimated gas mass fraction by this method is lower than that obtained above by a factor of up to $\sim1.3$ at $M_{*}\lesssim10^{10}$ M$_{\odot}$. The data point in the left panel of Fig. \ref{fig:ObservedDataPoints} at this stellar mass range would be lower if the estimation from the main-sequence is more accurate. If we do not use the data point in the lowest mass bin in the following analysis, however, the main results in this work do not change significantly.

\begin{figure}[!htb]
\includegraphics[angle=270, scale=0.53]{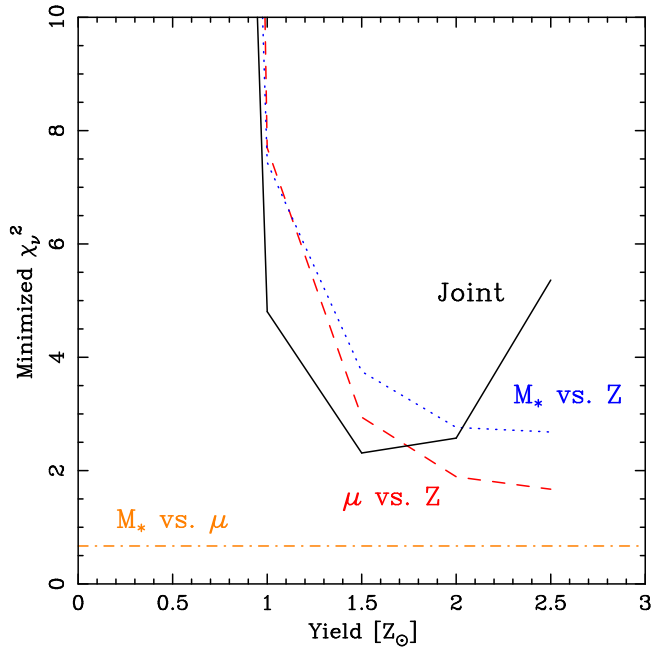}
\caption{The minimized reduced $\chi^{2}$ as a function of yield. The results of the fitting for $M_{*}$ vs. $\mu$, $\mu$ vs. $Z$, $M_{*}$ vs. $Z$ diagrams, and joint fitting are presented by \textit{dot-dashed}, \textit{dashed}, \textit{dotted}, and \textit{solid} line, respectively. \label{fig:VariousYield}}
\end{figure}

\subsection{Analytic Models with Gas Inflow and Outflow\label{sec:AnalyticModels}}
The observed data points are fitted by using simple analytic models with gas inflow and outflow. In a closed box model, the metallicity simply connected to the gas mass fraction as $Z=y_{eff}\textrm{ln}(1/\mu)$, where $Z$, $y_{eff}$, and $\mu$ is the metallicity, effective yield, and the gas mass fraction, respectively \citep{Pagel:1997p19846}. With the process including the gas flow, the formulation becomes somewhat complex. According to the notation by \citet{2001ASSL..253.....M}, we formulate the metallicity ($Z$) as a function of gas mass fraction ($\mu$) as below:

\begin{equation}
Z=\frac{y_{\scalebox{0.6}{Z}}}{f_{i}}\{1-[(f_{i}-f_{o})-(f_{i}-f_{o}-1)\mu^{-1}]^{\frac{f_{i}}{f_{i}-f_{o}-1}}\}, \label{eq:ModelZ}
\end{equation}
where $y_{\scalebox{0.6}{Z}}$ is a true yield, and $f_{i}$ and $f_{o}$ are infall and outflow rate which are normalized by the SFR, respectively, and are time-independent quantities \citep[see also][for the derivation]{Recchi:2008, Spitoni:2010p26344}. The outflow rate proportional to the SFR is realistic if the supernovae and massive stars are dominant drivers of the galactic wind. Although the detailed process of the gas accretion is poorly understood, the inflow rate proportional to the SFR would be reasonable because the amount of the infalling gas is closely related to the gas available to form stars. \citet{Recchi:2008} compare the similar simple analytical model with inflow rate proportional to the SFR and that with exponentially declining accretion with time, where the linear Schmidt law is assumed. They show that the difference between two models is not substantial, and the inflow rate is almost proportional to the SFR. Recent theoretical studies by \citet{Bouche:2010p12762} and \citet{Lilly:2013p26256} also show that the inflow rate is proportional to the SFR under some simple conditions. In the simple analytic model, we assume that the infalling gas is metal free and the metallicity of the outflowing gas is equal to that of the ISM. The latter assumption is reasonable for massive galaxies ($\gtrsim10^{9.5}$ M$_{\odot}$), where the local mass-metallicity relation is reproduced by the model with outflow whose metallicity is comparable to that of the ISM \citep{Spitoni:2010p26344}.
The true yield is defined as:
\begin{equation}
y_{\scalebox{0.6}{Z}}=\frac{1}{1-R}\int_{1}^{\infty}m p_{\scalebox{0.6}{Z}m} \phi(m) dm,
\end{equation}
 where $\phi(m)$ is the IMF, $R$ is the total mass fraction which is restored to the ISM or ``return fraction'', and $p_{\scalebox{0.6}{Z}m}$ is the fraction of newly produced and ejected metals by a star of mass $m$.  Although the true yield is highly uncertain, here we assume that $y_{\scalebox{0.6}{Z}}=1.5$ Z$_{\odot}$ in this work after previous works \citep[e.g.,][]{Erb:2008p4784}. We will discuss effects of varying the true yield on the results at the end of this section. The gas mass fraction can be written as a function of the stellar mass ($M_{*}$) as follows:
\begin{equation}
\mu = \frac{M^{0}_{gas}+(f_{i}-f_{o}-1)M_{*}}{M^{0}_{gas}+(f_{i}-f_{o})M_{*}}, \label{eq:ModelMu}
\end{equation}
where $M^{0}_{gas}$ is the initial primordial gas mass. The relation between metallicity and the stellar mass can be derived by substituting Eq. \ref{eq:ModelMu} into Eq. \ref{eq:ModelZ}. It is worth noting that the equilibrium condition, which will be mentioned in Sec. \ref{sec:CosmicEvolution}, occurs when $f_{i}-f_{o}-1=0$.

\def\arraystretch{1.5}
\begin{table*}
 \begin{center}
 \caption{The best-fit inflow rate and the outflow rate}\label{tab:results}
 \begin{tabular}{cccccccccc}
 \hline \hline
 redshift & \multicolumn{4}{c}{inflow rate ($f_{i}$)} & & \multicolumn{4}{c}{outflow rate ($f_{o}$)} \\
\cline{2-5}\cline{7-10}
 & $M_{*}$ vs. $\mu$ &  $\mu$ vs. $Z$ & $M_{*}$ vs. $Z$ & Joint & & $M_{*}$ vs. $\mu$ & $\mu$ vs. $Z$ & $M_{*}$ vs. $Z$ & Joint \\
\hline
0.0 & $2.92^{+0.06}_{-1.86}$ & $1.53^{+0.15}_{-0.20}$ & $1.57^{+0.12}_{-0.14}$ & $1.56^{+0.15}_{-0.12}$ & & $1.80^{+0.12}_{-1.80}$ & $0.64^{+0.66}_{-0.30}$ & $0.50^{+0.20}_{-0.50}$ & $0.44^{+0.16}_{-0.16}$\\
1.4 & $2.40^{+0.58}_{-1.24}$ & $1.67^{+0.21}_{-0.57}$ & $1.77^{+0.16}_{-0.10}$ & $1.84^{+0.14}_{-0.14}$ & & $1.16^{+0.66}_{-1.16}$ & $0.98^{+0.78}_{-0.46}$ & $0.18^{+0.92}_{-0.18}$ & $0.60^{+0.18}_{-0.22}$\\
2.2 & $2.62^{+0.36}_{-1.62}$ & $1.44^{+0.92}_{-1.44}$ & $2.02^{+0.36}_{-0.21}$ & $2.17^{+0.48}_{-0.36}$ & & $1.42^{+0.56}_{-1.42}$ & $2.58^{+1.40}_{-1.80}$ & $0.00^{+1.44}_{-0.00}$ & $0.94^{+0.58}_{-0.64}$\\
\hline
\end{tabular}
\end{center}
\end{table*}

\begin{figure*}[!htb]
\includegraphics[angle=270, scale=0.37]{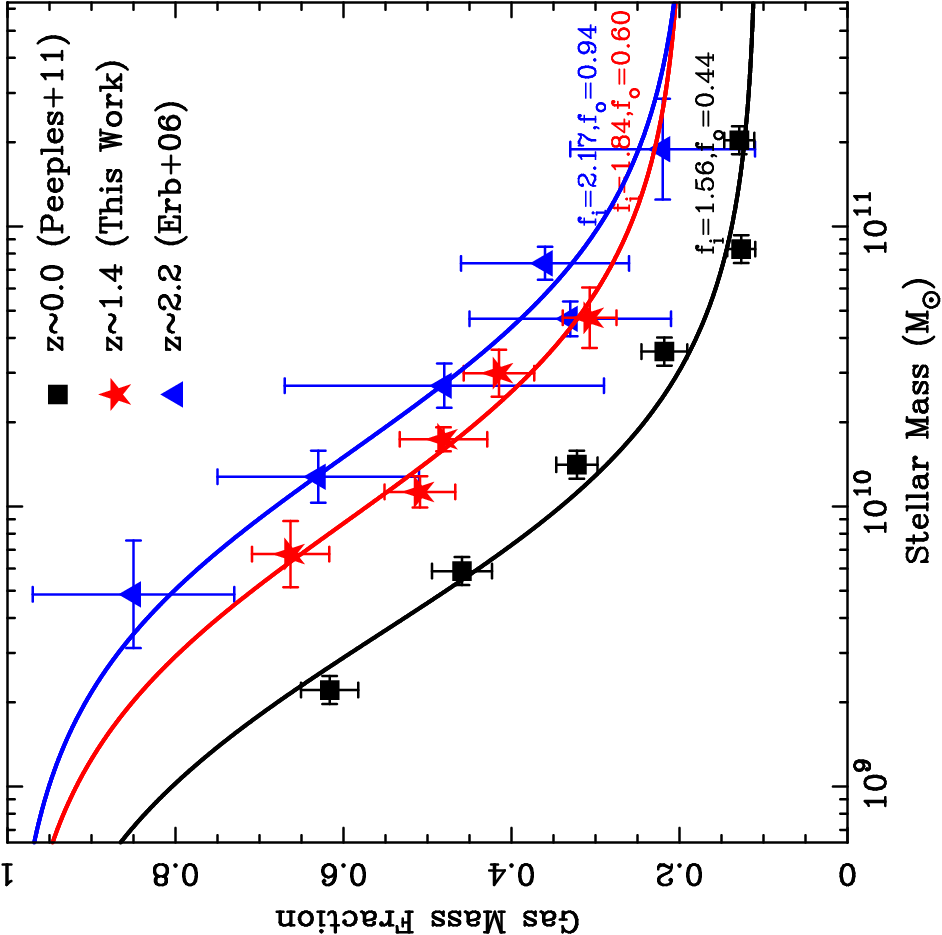}
\includegraphics[angle=270, scale=0.37]{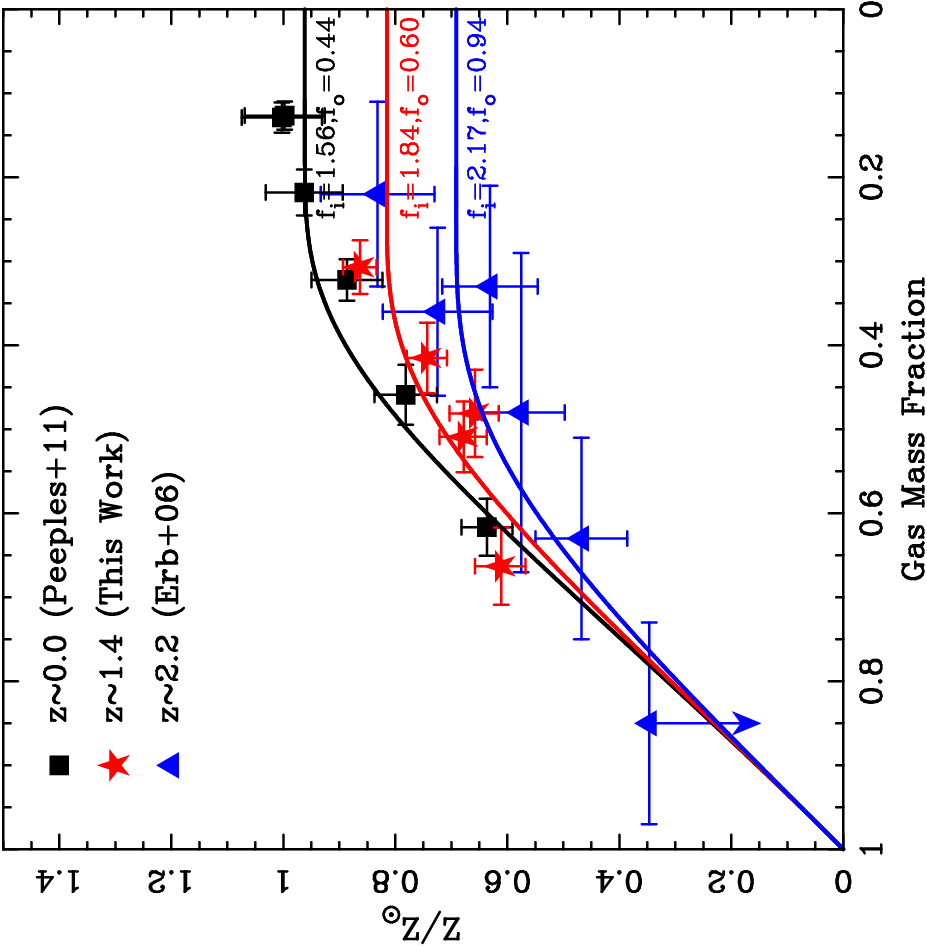}
\includegraphics[angle=270, scale=0.37]{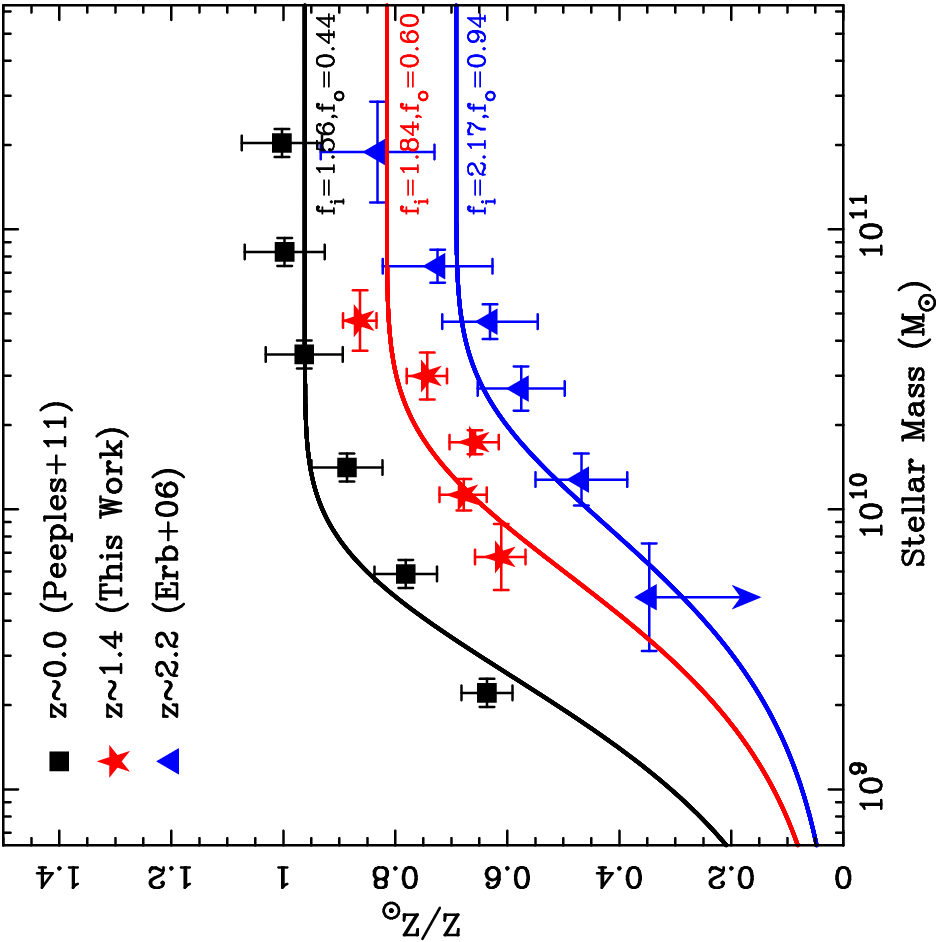}
\caption{The comparison of our result (\textit{filled stars}) to those at $z\sim0$ (\textit{filled squares}) and $z\sim2.2$ (\textit{filled triangles}) for the stellar mass vs. gas mass fraction ($M_{*}$ vs. $\mu$) diagram (\textit{left}), the gas mass fraction vs. metallicity ($\mu$ vs. $Z$) diagram (\textit{middle}), and stellar mass vs. metallicity ($M_{*}$ vs. $Z$) diagram (\textit{right}). We use data for the stellar mass, gas mass fraction, and the metallicity at $z\sim2.2$ by \citet{Erb:2006p4143}. For the data of the stellar mass and the gas mass fraction at $z\sim0$, we use the compiled result of star-forming galaxies at $z\sim0$ by \citet{Peeples:2011p22365}. The metallicity of the sample at $z\sim0$ is derived from the stellar mass by assuming the mass-metallicity relation. We use the mass-metallicity relation at $z\sim0.1$ by \citet{Erb:2006p4143}, which is derived from the sample by \citet{Tremonti:2004p4119} re-calculated by using the N2 method. The best-fit models from the joint fitting are shown by \textit{solid lines} for each redshift sample.\label{fig:Comparison}}
\end{figure*}

The average data points presented as the filled stars on the three diagrams (gas mass fraction vs. metallicity, stellar mass vs. gas mass fraction, and the stellar mass vs. metallicity) are fitted with the simple analytics models, taking the inflow rate, the outflow rate as a free parameter. The parameter range of the inflow rate and outflow rate is $f_{i}=0.0-3.0$ and $f_{o}=0.0-4.0$, respectively, with 200 grids per each parameter. We then carry out the standard $\chi^{2}$ minimization fitting for the data points. As presented in Eq. \ref{eq:ModelMu}, the initial gas mass can be a free parameter for the fitting of plots of stellar mass vs. gas mass fraction and the stellar mass vs. metallicity. Although the best-fit initial gas mass can be derived from the plots of stellar mass vs. gas mass fraction and the stellar mass vs. metallicity independently, we use the best-fit value of $M^{0}_{gas}=10^{10.04}$ M$_{\odot}$ derived from the stellar mass vs. gas mass fraction by varying in the range of $M^{0}_{gas}=10^{9} - 10^{11}$ M$_{\odot}$ with 50 grids, and fix the initial gas mass to this value in the fitting of the stellar mass vs. metallicity plot. If we take the initial gas mass as a free parameter in the stellar mass vs. metallicity fitting, the best fit initial gas mass is $10^{9.92}$ M$_{\odot}$, and the effects on the resulting inflow and outflow rate are very small ($<0.05$). In Eq. \ref{eq:ModelZ}, the term $(f_{i}-f_{o})-(f_{i}-f_{o}-1)\mu^{-1}$ must be positive. Due to this constraint condition, the range of gas mass fraction is defined for a given inflow rate and outflow rate, as pointed out previously \citep{Recchi:2008, Spitoni:2010p26344}. In this work, only models in which all of the observed gas mass fractions can be in the available range are used in the fitting process. The best-fit models are indicated as \textit{red dashed lines} in Fig. \ref{fig:ObservedDataPoints}.

Fig. \ref{fig:Chisq:z14} shows the $\chi^{2}$ maps with $1\sigma$, $2\sigma$, and $3\sigma$ contours against the inflow rate and the outflow rate for each fitting. In the top left panel of Fig. \ref{fig:Chisq:z14}, a severe degeneracy between the inflow rate and outflow rate in the fitting for the stellar mass vs. gas mass fraction is shown; the lower inflow rate and the lower outflow rate show a similar $\chi^{2}$ value as the case of the higher inflow rate and the higher outflow rate. Because the gas mass fraction decreases with increasing outflow rate at a fixed stellar mass, the higher inflow rate, which increases the gas mass fraction at a fixed  stellar mass, is needed to reproduce the observed data points. The best-fit inflow rate and the outflow rate are $2.40^{+0.58}_{-1.24}$ and $1.16^{+0.66}_{-1.16}$, respectively. In the top right panel of Fig. \ref{fig:Chisq:z14}, it is shown that degeneracy between the inflow rate and the outflow in the fitting for plot of the gas mass fraction vs. metallicity; the lower inflow rate and the higher outflow rate show a similar $\chi^{2}$ value as the higher inflow rate and the lower outflow rate. The best-fit inflow rate and the outflow rate are $1.67^{+0.21}_{-0.57}$ and $0.98^{+0.78}_{-0.46}$, respectively. The sharp cut-off in the contour around $f_{i}\sim1.9$ and $f_{o}\sim0.5$ arises from the constraint condition in Eq. 1 as described above. In the bottom left panel of Fig. \ref{fig:Chisq:z14}, the $\chi^{2}$ map for the plot of the stellar mass vs. metallicity is presented. There exists no severe degeneracy between inflow rate and outflow rate unlike the former two cases. The best-fit inflow rate and the outflow rate are $1.77^{+0.16}_{-0.10}$ and $0.18^{+0.92}_{-0.18}$, respectively. As with the case with the top right panel of Fig. \ref{fig:Chisq:z14}, the sharp cut-off in the contour around $f_{i}\sim1.9$ and $f_{o}\sim1.1$ also arises from the constraint condition in Eq. 1. 

Although there exists degeneracy more or less in each fitting, the direction of the degeneracy are basically different. The combined fitting would be, therefore, effective for the constraint of the parameters. We define the joint $\chi^{2}$ as $\chi^{2}_{Joint} \equiv \chi^{2}_{M_{*}\ vs.\ \mu} + \chi^{2}_{M_{*}\ vs.\ Z}$. Here, we use the stellar mass, which is the most robust quantity among the three observables, as an independent variable\footnote{In the joint fitting, all the other combinations, such as $\chi^{2}_{Joint} \equiv \chi^{2}_{M_{*}\ vs.\ \mu} + \chi^{2}_{\mu\ vs.\ Z}$, $\chi^{2}_{Joint} \equiv \chi^{2}_{\mu\ vs.\ Z} + \chi^{2}_{M_{*}\ vs.\ Z}$, and $\chi^{2}_{Joint} \equiv \chi^{2}_{M_{*}\ vs.\ \mu} + \chi^{2}_{\mu\ vs.\ Z} + \chi^{2}_{M_{*}\ vs. \ Z}$ show almost the same best-fit inflow and outflow rate.}. In the bottom right panel of Fig. \ref{fig:Chisq:z14}, the joint $\chi^{2}$ map is presented. The least-$\chi^{2}$ value can be found at the inflow rate of $1.84^{+0.14}_{-0.14}$ and the outflow rate of $0.60^{+0.18}_{-0.22}$. The resulting best-fit inflow and outflow rate are also summarized in Tab. \ref{tab:results}. The best-fit models from the joint fitting are also presented as \textit{black solid lines} in Fig. \ref{fig:ObservedDataPoints}.

Since the true yield of high redshift galaxies is highly uncertain, we assume the fixed yield of 1.5 Z$_{\odot}$ in Eq. \ref{eq:ModelZ} to reduce the number of the free parameters. We examine, here, the effect of the yield on the above results, by varying the yield. The observed data of the stellar mass vs. gas mass fraction, gas mass fraction vs. metallicity, and the stellar mass vs. metallicity are fitted in the same manner as presented above but with using different yield fixed at $y_{\scalebox{0.6}{Z}}=0.5$, 1.0, 2.0, and 2.5 Z$_{\odot}$. Both of the best-fit inflow and outflow rate increase with increasing yield. The best fit inflow rates are $f_{i}=0.81$ to $2.97$ in the case of $y_{\scalebox{0.6}{Z}}=0.5$ to $2.5$ Z$_{\odot}$, respectively, while the best fit outflow rates are $f_{o}=0.00$ to $1.72$ in the case of $y_{\scalebox{0.6}{Z}}=0.5$ to $2.5$ Z$_{\odot}$, respectively. In Fig. \ref{fig:VariousYield}, we show the minimum reduced chi-square value ($\chi_{\nu}^{2}$) in each case of yield. Because the yield does not affect the fitting for the diagram of $M_{*}$ vs. $\mu$, the minimum $\chi_{\nu}^{2}$ value is constant against the yield. Although the minimum $\chi_{\nu}^{2}$ for the case of $\mu$ vs. $Z$ and $M_{*}$ vs. $Z$ decreases with increasing yield, that for the joint fitting has a least value at $y_{\scalebox{0.6}{Z}}=1.5$ Z$_{\odot}$.

It may be realistic that the inflow and outflow nature depends on the mass of the galaxies. Here, we divide the sample into two sub-samples by the median total (gas $+$ stellar) mass of $3.4\times10^{10}$ M$_{\odot}$. The average total masses of the lower and higher total mass samples are $2.2\times10^{10}$ and $6.4\times10^{10}$ M$_{\odot}$, respectively. The same fitting procedure is done for each sub-sample. The best-fit initial gas mass is $M^{0}_{gas}=10^{10.08}$ M$_{\odot}$ and $M^{0}_{gas}=10^{10.36}$ M$_{\odot}$ for the sub-samples of lower and higher total masses, respectively. The obtained inflow rate of the sample with lower total mass and the higher total mass is $1.81^{+0.38}_{-0.44}$ and $1.72^{+0.15}_{-0.16}$, respectively. The outflow rate of the lower total mass sample and the higher total mass sample is $0.82^{+0.70}_{-0.82}$ and $0.70^{+0.36}_{-0.54}$, respectively. Both inflow rate and outflow rate agree with each other within the large errors. No clear stellar mass dependence of inflow rate and outflow rate at $M_{*}>10^{9.5}$ M$_{\odot}$ is also reported to reproduce the mass-metallicity relation at $z\sim0.1$ by \citet{Spitoni:2010p26344}.

\begin{figure*}[!htb]
\includegraphics[angle=270, scale=1.00]{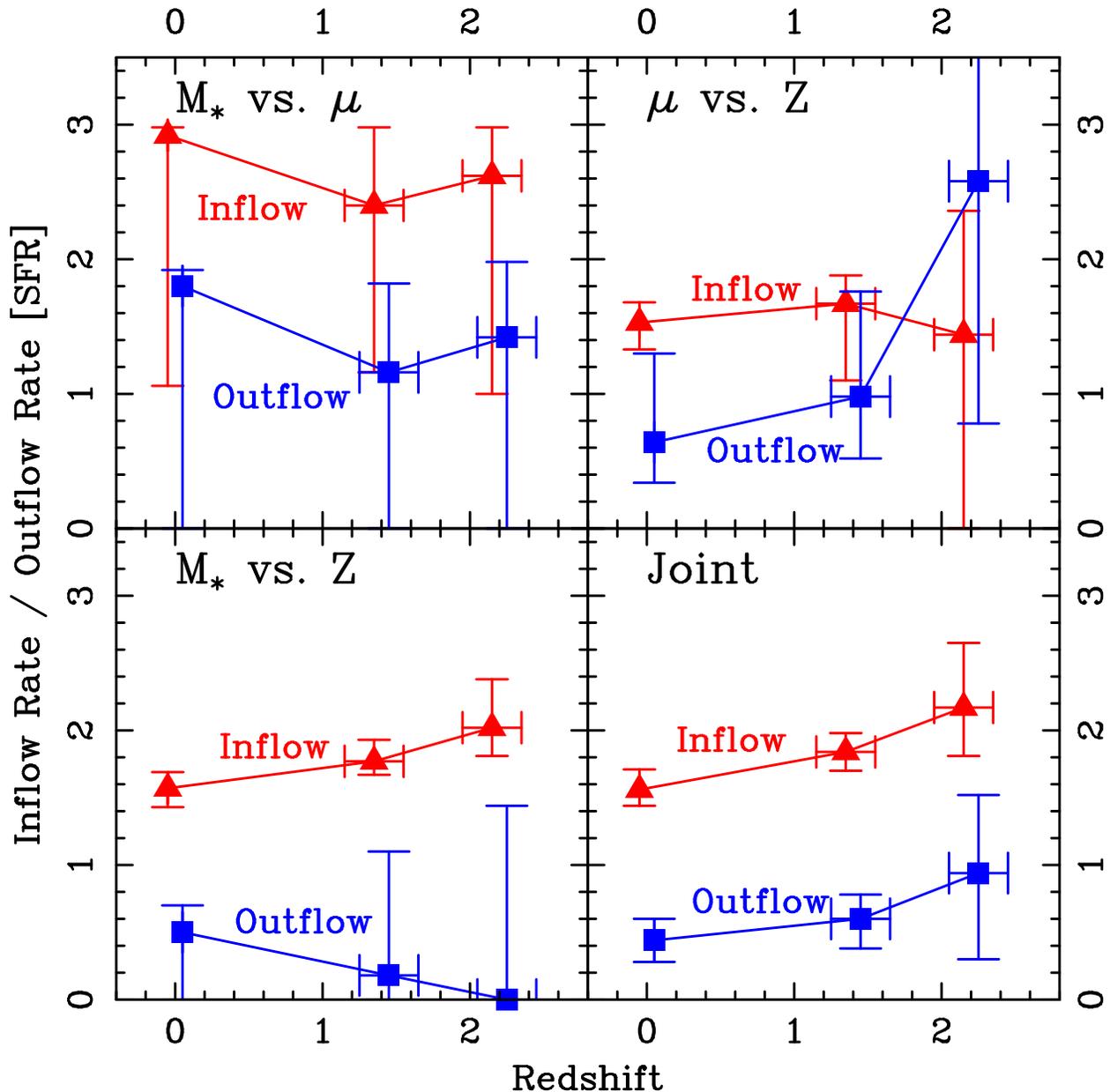}
\caption{Cosmic Evolution of the inflow rate (\textit{filled triangles}) and the outflow rate (\textit{filled squares}) in unit of SFR. The inflow / outflow rates are the best-fit values for the stellar mass vs. gas mass fraction ($M_{*}$ vs. $\mu$) diagram (\textit{top left}), the gas mass fraction vs. metallicity ($\mu$ vs. $Z$) diagram (\textit{top right}), stellar mass vs. metallicity ($M_{*}$ vs. $Z$) diagram (\textit{bottom left}), and the results from the joint fitting (\textit{bottom right}). The data points are slightly shifted ($\pm 0.05$) along the X-axis for clarity. \label{fig:RateEvolution}}
\end{figure*}

\begin{figure*}[!htb]
\includegraphics[angle=270, scale=0.50]{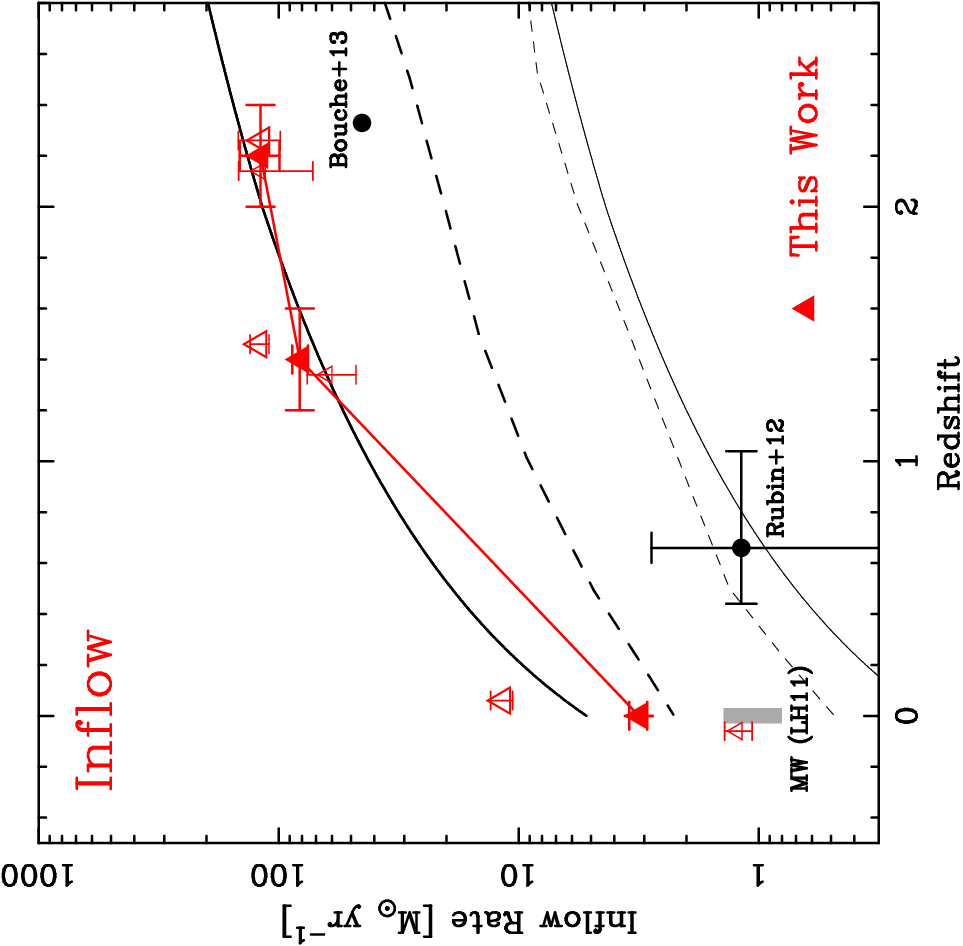}
\includegraphics[angle=270, scale=0.50]{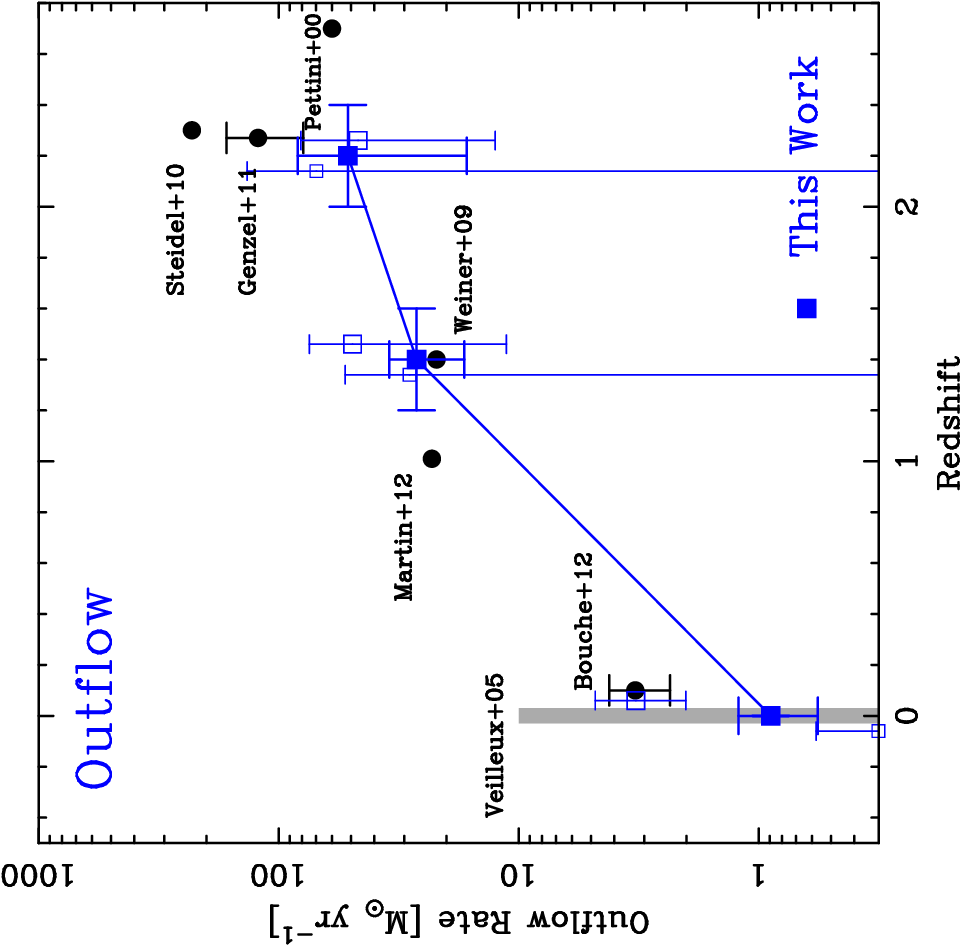}
\caption{\textbf{Left}: The inflow rate in unit of M$_{\odot}$ yr$^{-1}$ as a function of redshift derived from the joint fitting as presented in Fig. \ref{fig:RateEvolution}. The data points in this work are indicated as \textit{filled triangles}. The observed results by \citet{Rubin:2012p27422}, and \citet{Bouche:2013p26633} are indicated by \textit{filled circles}. The range of inflow rate of Milky Way obtained from ionized high-velocity clouds by \citet{Lehner:2011p29628} is indicated by the vertical \textit{gray shaded bar}. The theoretical expectations from the cosmological simulation by \citet{Keres:2005p26220} are presented in the cases of the dark halo mass at the redshift of $M_{DH}=10^{11}$ (\textit{thin dashed}) and $10^{12}$ M$_{\odot}$ (\textit{thick dashed}), while the analytic models by \citet{Bouche:2010p12762} are presented with the dark halo mass at the redshift of $M_{DH}=10^{11}$ (\textit{thin solid}) and $2\times10^{12}$ M$_{\odot}$ (\textit{thick solid}). The result of lower (higher) mass sub-sample is shown by smaller (larger) \textit{open triangles}, where the data points are shifted along x-axis by $-0.06$ ($+0.06$) for a clear display (see text for details). \textbf{Right}: The outflow rate in unit of M$_{\odot}$ yr$^{-1}$ as a function of redshift derived from the joint fitting as presented in Fig. \ref{fig:RateEvolution}. The data points in this work are shown by \textit{filled squares}. The range of the outflow rate of local galaxies by \citet{Veilleux:2005p14235} is shown by the vertical \textit{gray shaded bar}. The previous results by \citet{Bouche:2012p27878}, \citet{Martin:2012p27158}, \citet{Weiner:2009p13953}, \citet{Genzel:2011p27874}, \citet{Steidel:2010p10312}, and \citet{Pettini:2000p13759} are indicated by \textit{filled circles}. The result of lower (higher) mass sub-sample is shown by smaller (larger) \textit{open squares}, where the data points are also shifted along x-axis by $-0.06$ ($+0.06$).\label{fig:RateEvolutionComparison}}
\end{figure*}

\subsection{Cosmic Evolution of Gas Inflow Rate and Outflow Rate\label{sec:CosmicEvolution}}
The results of our sample at $z\sim1.4$ are compared to previous results at $z\sim2.2$ and $z\sim0$ in the plots of stellar mass vs. gas mass fraction, gas mass fraction vs. metallicity, and stellar mass vs. metallicity. We use data for the stellar mass, gas mass fraction, and the metallicity at $z\sim2.2$ by \citet{Erb:2006p4143}. The least massive (with the lowest metallicity) data point is not used in the fitting, because the obtained metallicity is an upper limit. As we mentioned in Section \ref{sec:GasMassFraction}, the gas mass fraction is limited due to the selection effect for the SFR limit sample. Since the sample at $z\sim2.2$ has a similar limiting SFR of $\sim10$ M$_{\odot}$ yr$^{-1}$ as our sample at $z\sim1.4$, the observable range of the gas mass fraction is almost the same. For the data of the stellar mass and the gas mass fraction at $z\sim0$, we use the results by \citet{Peeples:2011p22365}. The gas mass fraction for the sample at $z\sim0$ is derived from the direct observations of H$_{2}$ and \ion{H}{1} gas mass. It should be noted that the derivation of the gas mass fraction is different from that for the sample at $z\sim1.4$ and $z\sim2.2$. The un-observable area on the diagram of stellar mass vs. gas mass fraction is sufficiently lower than those of the sample at $z\sim1.4$ (gray shaded region in Fig. \ref{fig:ObservedDataPoints}) and $z\sim2.2$. The metallicity of these sample are derived from the stellar mass by assuming the mass-metallicity relation. In order to compare with the low-z sample, here we use that by \citet{Erb:2006p4143}, which is re-calculated by using the N2 method for the SDSS sample at $z\sim0.1$ by \citet{Tremonti:2004p4119}. Although this indicator saturates near solar metallicity \citep{Pettini:2004p7356}, the global trend would not be much affected.

Fig. \ref{fig:Comparison} shows the comparison of our result to those at other redshifts in the plots of stellar mass vs. gas mass fraction, gas mass fraction vs. metallicity, stellar mass vs. metallicity, respectively. With decreasing redshift from $z\sim2.2$ to $z\sim0$, the gas mass fraction decreases at a fixed  stellar mass, the metallicity increases at a fixed gas mass fraction and stellar mass. For these sample at various redshifts, the same fitting method by using the simple analytic models is applied. We obtained similar $\chi^{2}$ maps for sample at $z\sim0$ and $z\sim2.2$ to that of our sample at $z\sim1.4$. Based on the same joint fitting as $z\sim1.4$, the best-fit inflow rate and outflow rate at $z\sim0$ are $1.56^{+0.15}_{-0.12}$ and $0.44^{+0.16}_{-0.16}$, respectively, and the best-fit inflow rate and outflow rate at $z\sim2.2$ are $2.17^{+0.48}_{-0.36}$ and $0.94^{+0.58}_{-0.64}$, respectively. For each plot, the best-fit models are presented by solid lines from $z\sim2.2$ to $z\sim0$ in Fig. \ref{fig:Comparison}. The results are also summarized in Tab. \ref{tab:results}.

In Fig. \ref{fig:RateEvolution}, the resulting inflow and outflow rate are presented as a function of redshift. In the fitting for stellar mass vs. gas mass fraction (top left panel), both inflow rate and outflow rate decrease from $z\sim2.2$ to $z\sim1.4$ and increase from $z\sim1.4$ to $z\sim0$. In the fitting for gas mass fraction vs. metallicity (top right panel), the inflow rate is almost constant from $z\sim2.2$ to $z\sim0$, while the outflow rate increases with increasing redshift. In the fitting of the stellar mass vs. metallicity (bottom left panel), the inflow rate increases, while the outflow rate decrease with increasing redshift. However, there exists large uncertainty in some cases due to the severe degeneracy. In the joint fitting (bottom right panel), both inflow rate and outflow rate increase with increasing redshift, suggesting higher activity of gas flow at higher redshift. It is worth noting that the resulting inflow rate and outflow rate show that the equilibrium condition, i.e., $f_{i}-f_{o}-1=0$ \citep[e.g.,][]{Dave:2011a}, is broadly satisfied.

The resulting inflow and outflow rate derived from the joint fitting as a function of redshift are compared to the results in the previous works with different methods from ours. In Fig. \ref{fig:RateEvolution}, the obtained rates of the gas inflow and outflow are presented as the normalized value by the SFR at the redshift. In order to compare with various observational results, we converted these values into the rate in unit of M$_{\odot}$ yr$^{-1}$. The actual inflow and outflow rate are represented as $f_{i} (1-R) \textrm{SFR}$ and $f_{o} (1-R) \textrm{SFR}$. The return fraction ranges from $\sim0.2$ to $\sim0.5$ depending on the IMF, which results in some uncertainty regarding the inflow and outflow rate. Here, we assumed the return fraction for the Salpeter IMF of $R=0.27$ \citep{Madau:2014}. We use the average SFRs of $59.2$ and $73.1$ M$_{\odot}$ yr$^{-1}$ for the sample at $z\sim1.4$ and $z\sim2.2$, respectively. For the sample at $z\sim0$, we estimate the typical SFR from the average stellar mass by using the relation between the stellar mass and the sSFR \citep[see][]{Peeples:2011p22365}. The resulting SFR of the sample at $z\sim0$ is $2.7$ M$_{\odot}$ yr$^{-1}$. Thus, the inflow rate at $z\sim0$, 1.4, and 2.2 are $3.2_{-0.2}^{+0.3}$, $81.7_{-6.2}^{+6.2}$, and $119.0_{-19.7}^{+26.3}$ M$_{\odot}$ yr$^{-1}$, respectively. In the left panel of Fig. \ref{fig:RateEvolutionComparison}, the obtained results are presented with the previous observational result for Milky Way by \citet{Lehner:2011p29628}, the result by \citet{Rubin:2012p27422} at $z\sim0.7$ and that by \citet{Bouche:2013p26633} at $z\sim2.3$, which show lower inflow rates than the result in this work.

In the left panel of Fig. \ref{fig:RateEvolutionComparison}, we present the comparison to the theoretical predictions from the cosmological simulations by \citet{Keres:2005p26220} in the cases of the dark matter halo mass ($M_{DH}$) at the redshift of $10^{11}$ and $10^{12}$ M$_{\odot}$. We also present the results from the analytic models by \citet{Bouche:2010p12762} with $M_{DH}$ at the redshift of 10$^{11}$ and $2\times10^{12}$ M$_{\odot}$, which correspond to the minimum and the maximum threshold in their model, respectively. In the comparison to the analytic models, we use Eq. 6 by \citet{Bouche:2010p12762} with the baryonic fraction of 0.18 the accretion efficiency of $\epsilon_{in}(z)=0.70 f(z)$. Here, we use a declining function $f(z)$ linear in time with the boundary conditions $f(z=2.2)=1$ and $f(z=0)=0.5$. The decline of the inflow rate from $z\la2$ is expected from the evolution of the linear growth factor \citep{Bouche:2010p12762}. As presented in the left panel of Fig. \ref{fig:RateEvolutionComparison}, the global trend of decreasing inflow rate from $z\sim2.2$ to $z\sim0$ agrees with the both theoretical models. The absolute inflow rates obtained in this work at $z\sim2.2$ to $z\sim0$ are generally close to the analytic model prediction by \citet{Bouche:2010p12762} with $M_{DH}=2\times10^{12}$ M$_{\odot}$, though they show larger absolute inflow rates than the cosmological simulations by \citet{Keres:2005p26220} with $M_{DH}=10^{11}$ and $10^{12}$ M$_{\odot}$. It should be noted that the typical gas mass plus stellar mass of our sample is $\sim4\times10^{10}$ M$_{\odot}$ and the ratio to the dark matter halo mass is $\sim0.02$ for  $M_{DH}$ of $2\times10^{12}$ M$_{\odot}$. Such low ratio is consistent with the models by \citet{Bouche:2010p12762}.

In the right panel of Fig. \ref{fig:RateEvolutionComparison}, the resulting outflow rates at $z\sim0$, 1.4, and 2.2 are presented in the comparison with previous observational results by \citet{Veilleux:2005p14235} and \citet{Bouche:2012p27878} at $z\sim0$, \citet{Martin:2012p27158} and \citet{Weiner:2009p13953} at $z\sim1$, and \citet{Genzel:2011p27874}, \citet{Steidel:2010p10312}, and \citet{Pettini:2000p13759} at $z\sim2$. The resulting outflow rate in this work at $z\sim0$, 1.4, and 2.2 are $0.9_{-0.3}^{+0.3}$, $26.6_{-9.8}^{+8.0}$, and $51.5_{-35.1}^{+31.8}$ M$_{\odot}$ yr$^{-1}$, respectively, which agree with the results obtained in the previous observational works. The increase of the outflow rate at high redshift would be naturally expected, because more efficient galactic wind is expected in lower mass galaxies with shallower gravitational potential well and abundant energetic sources such as supernovae are also expected in galaxies at high redshift. It is worth noting that, in the analytic model we used, the inflow and outflow rate are assumed to be constant in time, and thus the values in Fig. \ref{fig:RateEvolutionComparison} are average ones before the target redshift. The actual inflow and outflow rate at the target redshift, therefore, may be smaller than those presented in Fig. \ref{fig:RateEvolutionComparison}.

As we discussed in Sec. \ref{sec:AnalyticModels}, the difference between $f_{i}$ and $f_{o}$ in the lower total (= stellar + gas) mass sample and those in the higher total mass sample at $z\sim1.4$ is marginal. Here, we calculate the mass rate of the inflow and outflow by using the average SFR for each sub-sample. The obtained inflow rates in lower and higher total mass samples are $62.9^{+13.2}_{-15.3}$ M$_{\odot}$ yr$^{-1}$ and $121.0^{+10.6}_{-11.3}$ M$_{\odot}$ yr$^{-1}$, respectively. The outflow rates in lower and higher total mass samples are $28.5^{+24.3}_{-28.5}$ M$_{\odot}$ yr$^{-1}$ and $49.3^{+25.3}_{-38.0}$ M$_{\odot}$ yr$^{-1}$, respectively. Since we do not know the total mass of individual galaxies of the other redshift samples, the total mass dependence of the inflow and outflow rate for the sample at the other redshifts cannot be examined by using a similar method. Alternatively, we examine the stellar mass dependence of the inflow and outflow rate. Here, for the sample at $z\sim0$ and $z\sim2.2$, we define the data points at the stellar mass lower (higher) than $3\times10^{10}$ M$_{\odot}$ as the lower (higher) stellar mass sample, and derive the inflow and outflow rate by using only these data points. The average stellar mass of the lower (higher) mass sub-sample at $z\sim0$ and $z\sim2.2$ are $0.6\ (8.5) \times10^{10}$ M$_{\odot}$ and $1.2\ (8.7) \times10^{10}$ M$_{\odot}$, respectively. In Fig. \ref{fig:RateEvolutionComparison}, the resulting inflow and outflow rate are shown as open triangles and squares, respectively. At $z\sim2.2$, no clear trend of the stellar mass dependence of both inflow and outflow rate can be seen. The mass dependence of the inflow rate at $z\lesssim1.4$, i.e., the inflow rate is higher in more massive galaxies, is roughly in agreement with the model prediction by \citet{Bouche:2010p12762}. The outflow rate of the more massive galaxies at $z\lesssim1.4$ appears to be higher than that of the less massive galaxies. The interpretation, however, is difficult because the more massive galaxies would be in the deeper gravity potential well, which impedes the ejection of the material by the galactic wind, while the more massive galaxies show the larger SFR, resulting the stronger galactic wind. It should also be noted that the error bars in the obtained outflow rate are generally large.

\section{Conclusions and Summary}
We try to constrain the inflow rate and the outflow rate of star-forming galaxies at $z\sim1.4$ by using a simple analytic model for the chemical evolution of galaxies. The sample is taken from a large NIR spectroscopic sample observed with Subaru/FMOS in the SXDS/UDS field. The gas metallicity is measured from the [\ion{N}{2}]$\lambda$6584/H$\alpha$ emission line ratio, and the gas mass is derived from the extinction corrected H$\alpha$ luminosity by assuming the Kennicutt-Schmidt law. We employ the least-$\chi^{2}$ fittings with the analytic model for the observed gas mass fraction, stellar mass, and metallicity. The joint $\chi^{2}$ fitting show the best-fit inflow rate of $\sim1.8$ and the outflow rate of $\sim0.6$ in unit of SFR. By applying the same analysis for previous results at $z\sim0$ and $z\sim2.2$, it is shown that the both inflow rate and outflow rate decrease with decreasing redshift, which implies the high activity of gas flow process at high redshift. The decreasing trend of the resulting inflow rate from $z\sim2.2$ to $z\sim0$ agrees with that seen in the previous observational works with different methods, though the absolute values are generally larger than the previous works. The resulting inflow rate in this work from $z\sim2.2$ to $z\sim0$ is close to the theoretical model by \citet{Bouche:2010p12762} with the dark matter mass at the redshift of $M_{DH}=2\times10^{12}$ M$_{\odot}$. The outflow rate and its evolution from $z\sim2.2$ to $z\sim0$ in this work agree well with the independent estimations in the previous observational works.

\section*{Acknowledgements}
We are grateful to the FMOS support astronomer Kentaro Aoki for his support during the observations. We also appreciate Soh Ikarashi, Kotaro Kohno, Kenta Matsuoka, and Tohru Nagao for sharing fibers in their FMOS observations. We thank the referee for very constructive comments. KY was financially supported by a Research Fellowship of the Japan Society for the Promotion of Science for Young Scientists. KO is supported by the Grant-in-Aid for Scientific Research (C) (24540230) from Japan Society for the Promotion of Science (JSPS). We acknowledge support for the FMOS instrument development from the UK Science and Technology Facilities Council (STFC). We would like to express our acknowledgement to the indigenous Hawaiian people for their understanding of the significant role of the summit of Mauna Kea in astronomical research.

{\it Facilities:} \facility{Subaru (FMOS, S-Cam)}, \facility{UKIRT (WFCAM)}, \facility{Spitzer (IRAC, MIPS)}, \facility{Herschel (PACS, SPIRE)}.

\end{document}